\documentclass[aps,prl,reprint,superscriptaddress,amsmath,amssymb,bibnotes,showkeys]{revtex4-1}

\usepackage{graphicx}% Include figure files
\usepackage{amsmath}
\usepackage{dcolumn}% Align table columns on decimal point
\usepackage{bm}% bold math
\usepackage{braket} %Dirac notation
\usepackage{bbold}
\usepackage{diagbox}
\usepackage{multirow}
\usepackage{lpic} %label figures
\usepackage{url}
\usepackage{array}
\usepackage{color}
\usepackage[
 colorlinks=true,
 urlcolor=blue,
 citecolor=blue,
 linkcolor=blue,
 bookmarks=false,
 pdfstartview={FitH},
]{hyperref}% add hypertext capabilities
\usepackage[capitalize]{cleveref}

\begin{document}

\onecolumngrid
\textbf{This is a preprint of an article published in the Proceedings of the National Academy of Sciences (PNAS). The final authenticated version is available online at: \url{https://doi.org/10.1073/pnas.2020585118}.}
\twocolumngrid

\hfill \break
\hfill \break

\title{Quadrupolar charge dynamics in the nonmagnetic FeSe$_{1-x}$S$_x$ superconductors}

\author{Weilu~Zhang}
\email{weiluzhang41@gmail.com}
\affiliation{Department of Physics $\&$ Astronomy, Rutgers University, Piscataway, NJ 08854}
\affiliation{Department of Engineering and Applied Sciences, Sophia University, Tokyo 102-8554, Japan}

\author{Shangfei~Wu}
\affiliation{Department of Physics $\&$ Astronomy, Rutgers University, Piscataway, NJ 08854}

\author{Shigeru~Kasahara}
\thanks{Present address: Research Institute for Interdisciplinary Science, Okayama University, Okayama 700-8530, Japan.}
\affiliation{Department of Physics, Kyoto University, Kyoto 606-8502, Japan}

\author{Takasada~Shibauchi}
\affiliation{Department of Advanced Materials Science, University of Tokyo, Kashiwa 277-8561, Japan}

\author{Yuji~Matsuda}
\affiliation{Department of Physics, Kyoto University, Kyoto 606-8502, Japan}

\author{Girsh~Blumberg}
\email{girsh@physics.rutgers.edu}
\affiliation{Department of Physics $\&$ Astronomy, Rutgers University, Piscataway, NJ 08854}
\affiliation{National Institute of Chemical Physics and Biophysics, 12618 Tallinn, Estonia}


\date{\today}

\begin{abstract}
We use polarization-resolved electronic Raman spectroscopy to study quadrupolar charge dynamics in a nonmagnetic FeSe$_{1-x}$S$_x$ superconductor.   
We observe two types of long-wavelength \textit{XY} symmetry excitations: 
1) a low-energy quasi-elastic scattering peak (QEP) and 
2) a broad electronic continuum with a maximum at 55~meV.
Below the tetragonal-to-orthorhombic structural transition at \textit{T\textsubscript{S}(x)}, a pseudogap suppression with temperature dependence reminiscent of the nematic order parameter develops in the \textit{XY} symmetry spectra of the electronic excitation continuum.       
The QEP exhibits critical enhancement upon cooling toward \textit{T\textsubscript{S}(x)}.   
The intensity of the QEP grows with increasing sulfur concentration \textit{x} and maximizes near critical concentration \textit{x\textsubscript{cr}} $\approx$ 0.16, while the pseudogap size decreases with the suppression of  \textit{T\textsubscript{S}(x)}. 
We interpret the development of the pseudogap in the quadrupole scattering channel as a manifestation of transition from the non-Fermi liquid regime, dominated by strong Pomeranchuk-like fluctuations giving rise to intense electronic continuum of excitations in the fourfold symmetric high-temperature phase, to the Fermi liquid regime in the broken-symmetry nematic phase where the quadrupole fluctuations are suppressed.
\end{abstract}


\maketitle


The iron-based superconductors (FeSCs) exhibit a complex phase diagram with multiple competing orders. 
For most of the FeSCs, an electronic nematic phase transition takes place at $T_S$, which is followed by a magnetic phase transition at $T_N$~\cite{Fradkin_Annurev2010,Stewart_RMP2011,Paglione_nphys2010,Fernandes_NatPhys2014}.
Superconductivity emerges in close proximity to the electronic nematic and the antiferromagnetic orders. 
The highest superconducting (SC) transition temperature $T_c$ often occurs when nematic and magnetic orders are fully suppressed but the orbital/charge or spin fluctuations remain strong~\cite{Fisher_Science2012,Gallais_PRL2013,Shibauchi2014,Bohmer_PRL2014,LuDai_Science2014,Matsuura2017}.
The relationship between these fluctuations and superconductivity has been the focus of intense research~\cite{Matsuda_Nature2012,Gallais_PRL2013,Gnezdilov_PRB2013,Fernandes_NatPhys2014,Baek_nmat2014,Bohmer_2015,Chubukov_PRX2016,Gallais_review,Thorsmolle_PRB2016,Massat_2016,Curro_PRL2016,Erez_PRX2016,Massat_2016,Chubukov_PRL2017,Hanaguri_2017,Licciardello2019_Nature,Hussey_2019,Hackl_Arxiv2017,Gallais2020,kreisel2020remarkable,Shibauchi2020,Lazarevi2020,coldea2020,Hashimoto2020}.

The family of FeSe superconductors is the simplest system to elucidate the origin of orbital and charge fluctuations because for these materials nematicity appears in the absence of magnetic order~\cite{Cava_PRL2009,Baek_nmat2014,Bohmer_PRL2015}. 
At the ambient pressure, a structural phase transition that breaks the fourfold rotational symmetry ($C_4$) takes place at $T_S=$ 90~K. 
Strong electronic quadrupole fluctuations involving the charge transfer between the degenerate Fe 3$d_{xz}$ and 3$d_{yz}$ orbitals, which contribute to most of the electronic density of states near $E_F$, have been observed above $T_S$~\cite{Hosoi_pnas2016,Tanatar_PRL2016,Massat_2016,Hackl_ncommun2019}.
The degeneracy of the $d_{xz}$ and $d_{yz}$ orbitals is lifted in the symmetry-broken phase~\cite{Watson_PRB2017,Coldea_review2017,Borisenko_SciRep2017}, where although the lattice is only weakly distorted, a prominent anisotropy of the electronic properties was detected~\cite{Feng_PRL2016,JCDavis_Science2017,Hanaguri_2017}. 
For single crystals, superconductivity emerges in the nematic phase at $T_c \approx 9$\,K~\cite{Cava_PRL2009}, while for FeSe monolayer films deposited on SrTiO$_3$ substrate the $T_c$ can be enhanced by almost an order of magnitude~\cite{Xue_CPL2012,Xue_nmat2015,Lee32,Kostin2018}.
An unusual orbital-selective SC pairing has been reported by angle-resolved photo-emission spectroscopy (ARPES) and quasiparticle interference (QPI) studies in bulk FeSe: The SC gap energy is large only at a specific region of the nematic Fermi surfaces with the Fe\,3$d_{yz}$ orbital characters~\cite{Feng_PRL2016,JCDavis_Science2017,Hanaguri_2017,Kostin2018}. 

%Fig1
%%%%%%%%%%%%%%%%%%%%%%%%%%%%%%%%%%%%,
 \begin{figure}[]
\includegraphics[width=\columnwidth]{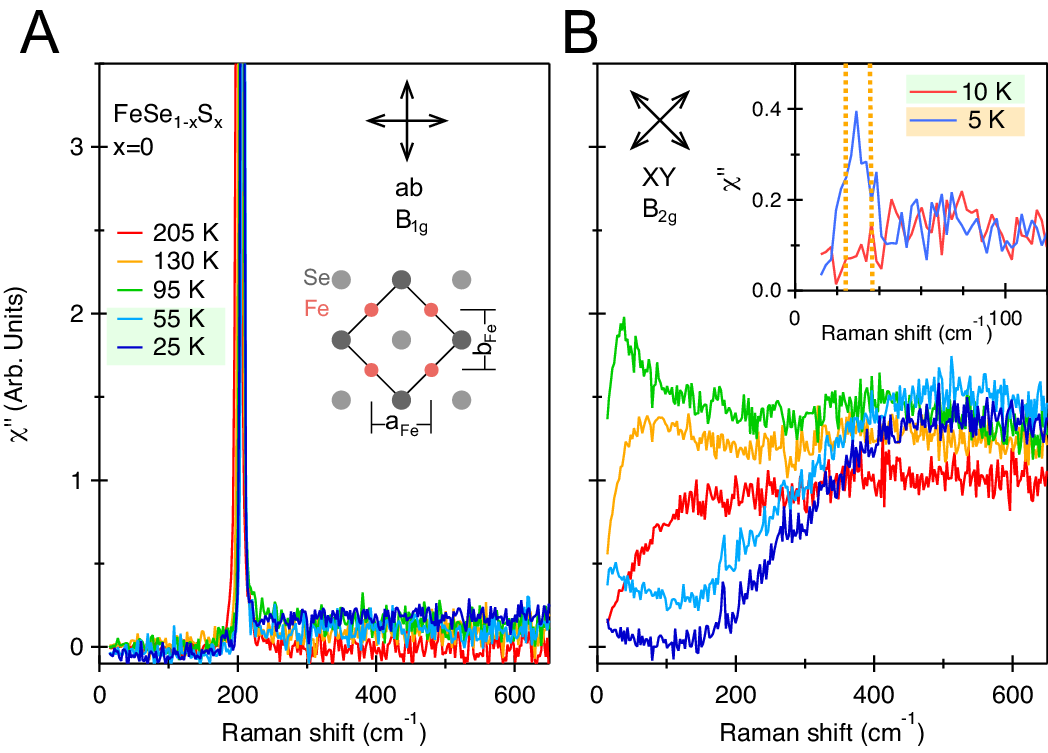}
 \caption{ \label{Fig1} 
 Temperature evolution of the $B_{1g}$($ab$) and $B_{2g}$($XY$) symmetry Raman response $\chi^{\prime\prime}(\omega,T)$ for undoped FeSe. A, \textit{Inset} shows the top view of the FeSe layer. 
 Dark and light gray circles represent the Se above and below the Fe layer. 
 The two-iron unit cell for the high-temperature phase is
 shown by solid lines. In the low-temperature phase, the nearest Fe-Fe
 bonding distance $a_{Fe}$ becomes larger than $b_{Fe}$ while $a_{Fe}$
 and $b_{Fe}$ remain orthogonal. B, \textit{Inset} shows $\chi^{\prime\prime}(\omega,T)$ in the $XY$ symmetry channel of FeSe in the normal state (red, 10 K) and the SC state (blue, 5 K). 
 The magnitude of the two SC gaps $2\Delta_{SC}$ = 3 and 4.6~meV measured by tunneling spectroscopy are shown with the vertical dotted lines (\textit{SI Appendix} and ref.~\cite{JCDavis_Science2017}).} 
 \end{figure}
%%%%%%%%%%%%%%%%%%%%%%%%%%%%%%%%%%%%%%

Partial isovalent sulfur substitution at the selenium site monotonically suppresses the structural phase transition temperature $T_S$ until it vanishes at the critical concentration $x_{cr} \approx 0.16$,  
while the SC transition temperature $T_c$ first mildly increases with substitution and reaches maximum value 11\,K at $x=0.08$~\cite{Hosoi_pnas2016,Hanaguri_2017,Licciardello2019_Nature}.
Thus, the phase diagram of FeSe$_{1-x}$S$_x$ alloys enables a spectroscopic study of the interplay between competing ordered phases. 

In this work we employ polarization-resolved Raman spectroscopy to study charge quadrupole dynamics in nonmagnetic superconductor alloy FeSe$_{1-x}$S$_x$~\cite{Hosoi_pnas2016,Matsuura2017}. 
We observe two main features in the $XY$ symmetry scattering channel:
1) a low-energy quasi-elastic scattering peak (QEP) that, above $T_S(x)$, exhibits enhancement and softening upon cooling in a wide temperature and sulfur doping range and 
2) a high-energy electronic continuum extending beyond 2,000~cm$^{-1}$ with a broad peak at 450~cm$^{-1}$ that arises due to beyond Fermi-liquid Pomeranchuk-like $XY$-quadrupole fluctuations in the high-temperature fourfold symmetric phase.
The Fermi-liquid regime recovers in the low-temperature nematic phase where the low-frequency quadrupole fluctuations are suppressed, causing an apparent pseudogap in the electronic continuum for metals with small but prominent Fermi surface pockets.

%Fig2,
%%%%%%%%%%%%%%%%%%%%%%%%%%%%%%%%%%%%
 \begin{figure*}[tbp]
%  \begin{figure}[tbp],
 \centering
 \includegraphics[width=2\columnwidth]{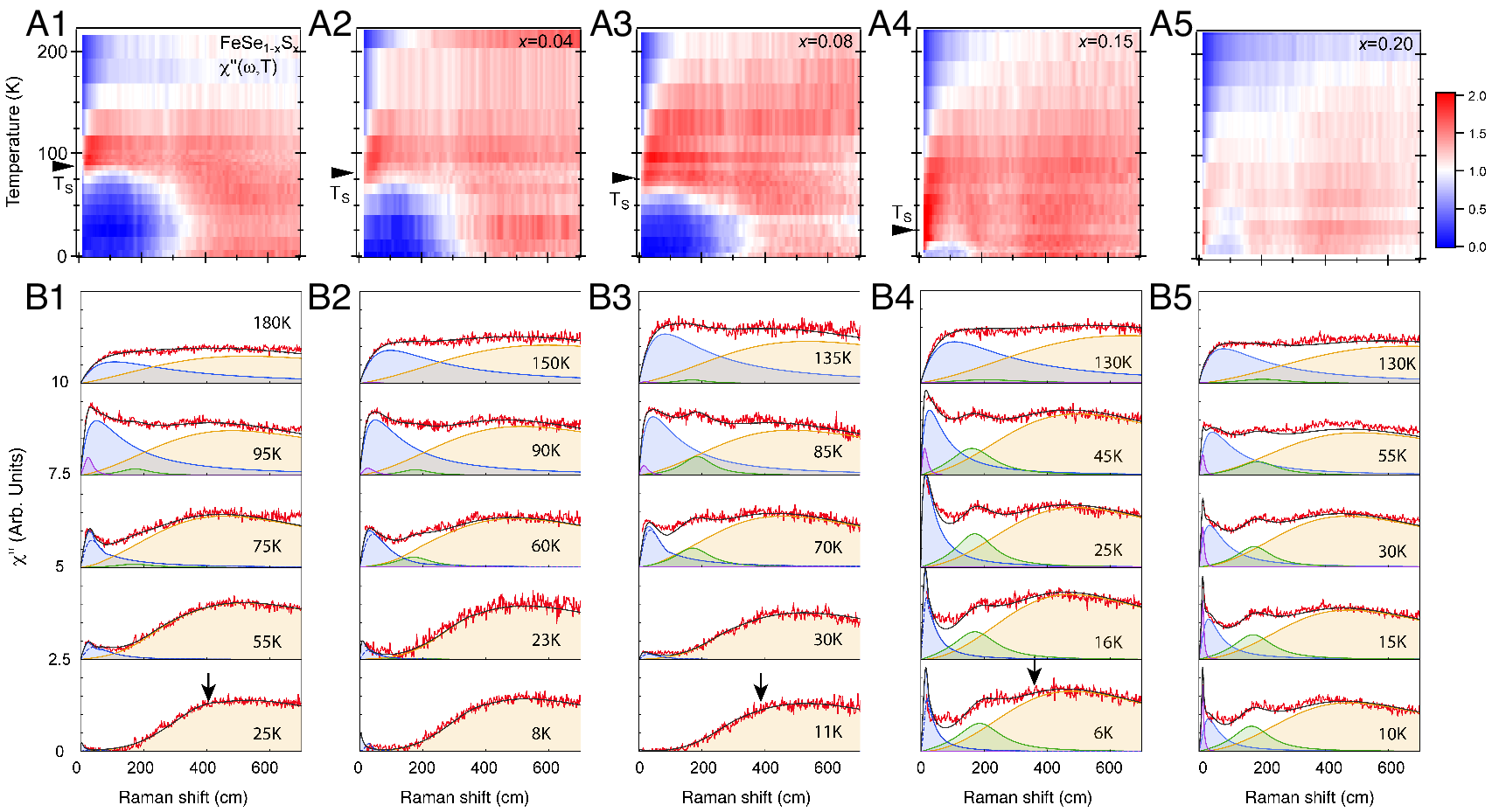}
 \caption{\label{Fig2} 
(A1-A5) 
Temperature evolution of Raman susceptibility $\chi_{XY}^{\prime\prime}(\omega, T)$ in the $XY$ symmetry channel for FeSe$_{1-x}$S$_x$ ($x$ = 0, 0.04, 0.08, 0.15 and 0.2). 
The arrows at the temperature axis denote $T_S(x)$.  
(B1-B5) 
$\chi_{XY}^{\prime\prime}(\omega, T)$ data (red) at representative temperatures and the fits (black) to the sum of the oscillators model.
The QEP contribution is shaded in blue; 
contribution of the strongly overdamped high-energy electronic oscillator is shaded in yellow. 
A feature due to interband transition at about 190\,cm$^{-1}$ is shown in green.
The additional low-frequency oscillator due to local lattice dynamics coupled to the fluctuating order parameter above $T_S(x)$~\cite{halperin1976,Billinge2019,Birgeneau2019} is shown in purple. 
Below $T_S(x)$, the coupled acoustic lattice mode and the QEP (shaded in blue) adds additional low-frequency spectral weight to the QEP (blue dashed lines). 
The arrows in B1, B3, and B4 indicate the energy of nematic $d_{xz}/d_{yz}$ orbital splitting reported in ARPES studies~\cite{Coldea_PRB2015,Yi_2019}. 
}
 \end{figure*}
%%%%%%%%%%%%%%%%%%%%%%%%%%%%%%%%%%%%%%

%Fig3
%%%%%%%%%%%%%%%%%%%%%%%%%%%%%%%%%%%%
\begin{figure}[!b]
\centering
\includegraphics[width=\columnwidth]{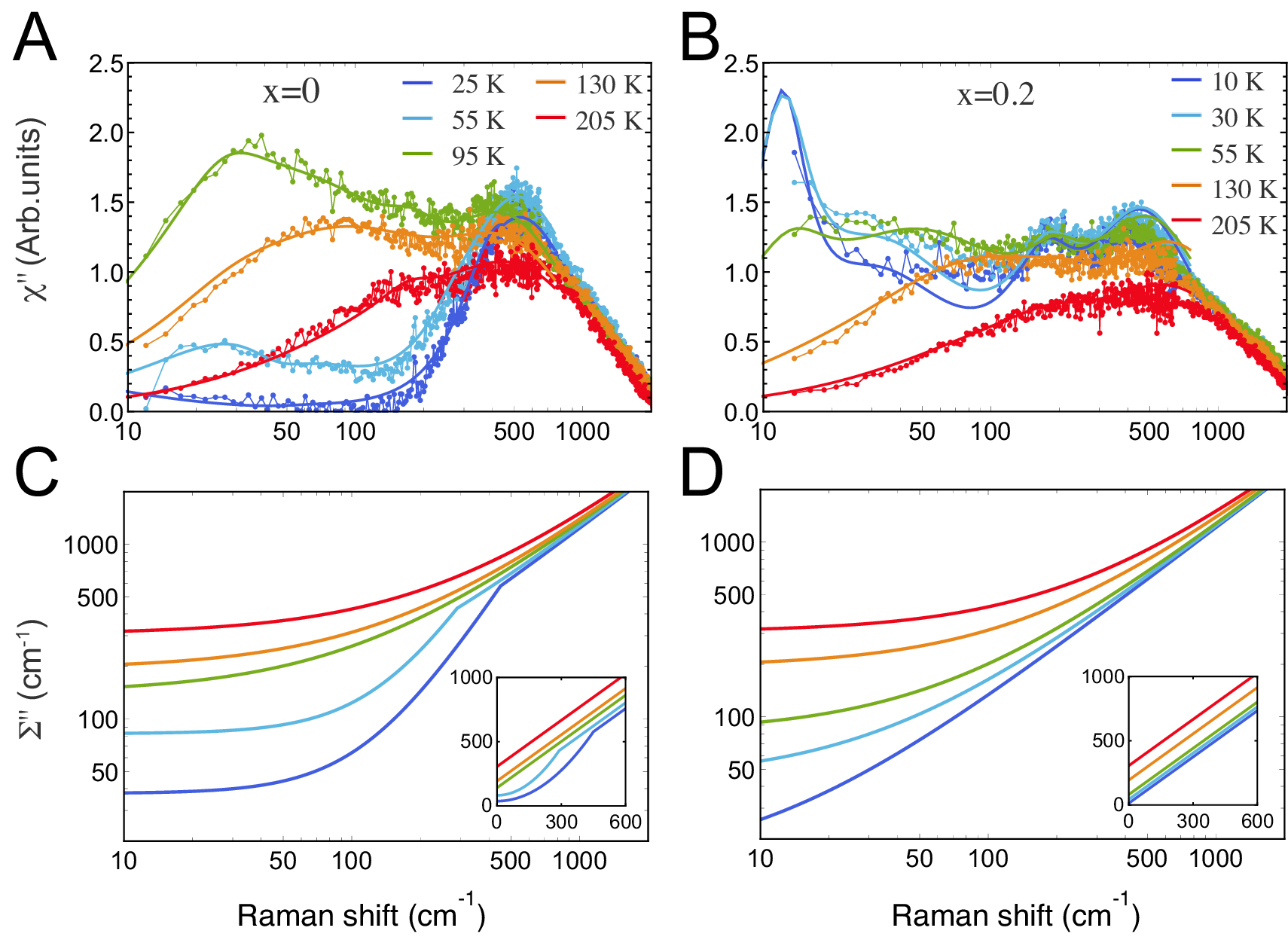}
\caption{\label{Fig3} 
(A and B) Temperature evolution of the $\chi_{XY}^{\prime\prime} (\omega,T)$ Raman response data and the fits to a model of oscillators (\textit{SI Appendix, Data Fit}) for stoichimetric FeSe in A and for FeSe$_{0.8}$S$_{0.2}$ in B. 
(C and D) The imaginary part of self-energy $\Sigma^{\prime\prime} (\omega,T)$ used for the data fits in A and B correspondingly. 
\textit{Insets} show a zoom-in of the low-frequency region for $\Sigma^{\prime\prime}(\omega,T)$ in linear scale.}
\end{figure}
%%%%%%%%%%%%%%%%%%%%%%%%%%%%%%%%%%%%%%

%Fig4
%%%%%%%%%%%%%%%%%%%%%%%%%%%%%%%%%%%%{}
\begin{figure*}[t]
\centering
\includegraphics[width=1.6\columnwidth]{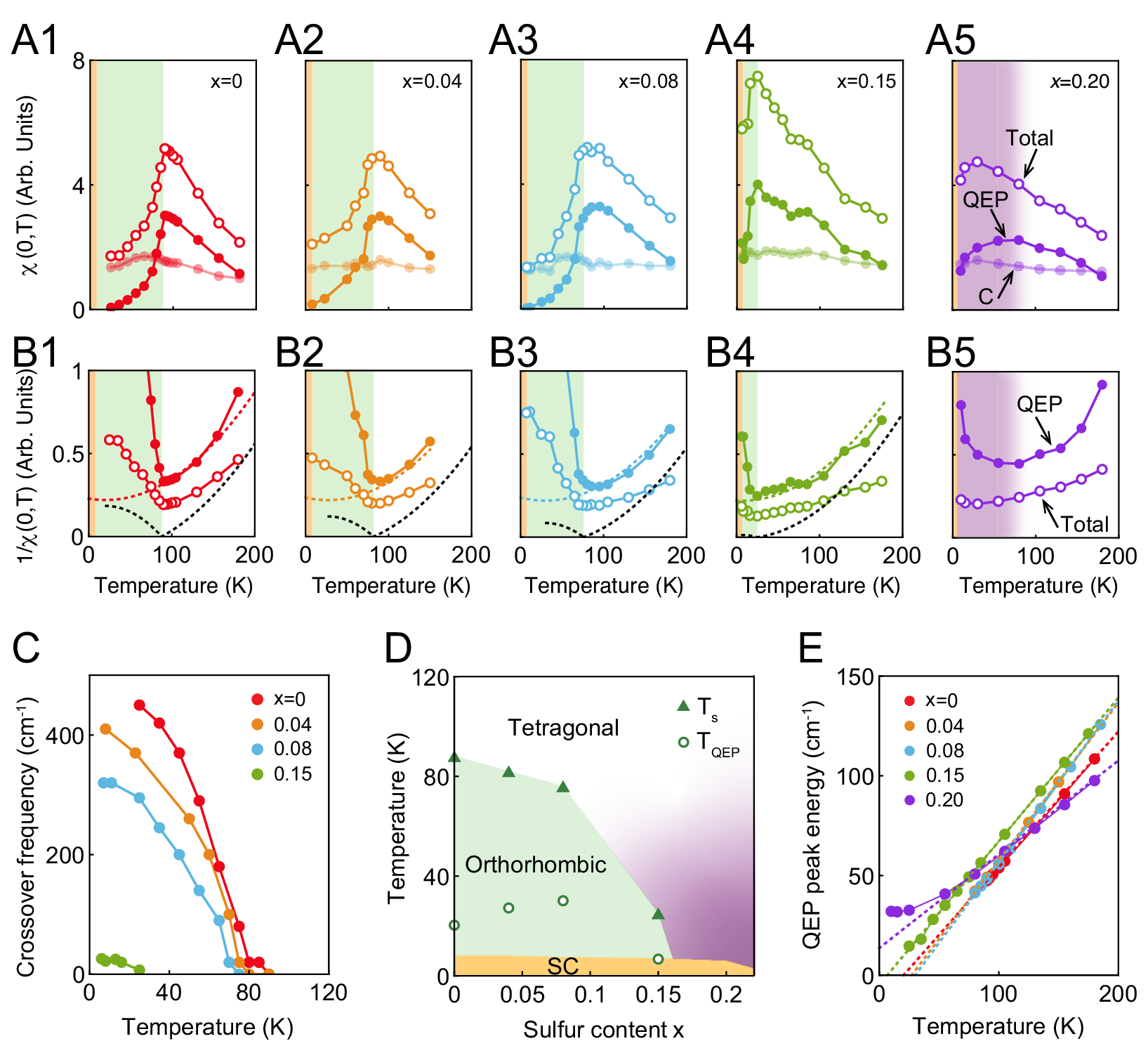}
\caption{\label{Fig4} 
(A1-A5)
Two main contributions to the $XY$-symmetry static Raman susceptibility: $\chi_{QEP}(0,T)$ [solid circles] and $\chi_{C}(0,T)$ [shaded circles] for $x$ = 0, 0.04, 0.08, 0.15 and 0.2. 
The open circles denote the total Raman static susceptibility (including all spectral features) obtained from the Raman data (Fig.\,\ref{Fig2}).
The green and yellow shades indicate temperature regions below $T_S(x)$ and $T_c(x)$ as shown in the phase diagram D~\cite{Hosoi_pnas2016}.
The purple shading in A5 and B5 denotes a region of phase diagram where the low-frequency fluctuations are significant (Fig.\,\ref{Fig2} and \textit{SI Appendix, Data Fit, section E})
(B1-B5) Temperature dependence of the inverse static Raman susceptibilities shown in A1-A5. The dashed color lines show the fit to parabolic function Eq.\,[\ref{fT}], and 
dashed black lines denote inverse total static susceptibilities including the lattice contribution, Eq.\,[\ref{totalchi}]. 
(C) Temperature evolution of the cross-over to Fermi-liquid boundary $\omega_c(\omega,T)$. 
Color coding for the respective sulfur concentration $x$ is same as in A1-A5.
(D) The temperature-sulfur concentration phase diagram.  
$T_S(x)$ is the nematic phase transition temperature~\cite{Hosoi_pnas2016};  
$T_{QEP}(x)$ is the temperature where QEP mode's peak frequency would soften to zero, as determined by linear approximation from the high-temperature phase (E).  
(E) Temperature dependence of the QEP peak frequency. 
Dashed asymptotic lines define $T_{QEP}(x)$. 
}
\end{figure*}
%%%%%%%%%%%%%%%%%%%%%%%%%%%%%%%%%%%%%%

\section*{Results}~~~
In Fig.\,\ref{Fig1}A and B we show temperature dependence of the Raman response for undoped FeSe in $B_{1g}$ ($ab$) and $B_{2g}$ ($XY$) symmetry channels ($D_{4h}$ point group) defined for a two-iron unit cell.
The data for the $B_{1g}$ channel are composed of the Fe phonon mode at 195~cm$^{-1}$~\cite{Gnezdilov_PRB2013} above a weak temperature independent continuum background( Fig.\,\ref{Fig1}A). 
In contrast, the electronic Raman continuum in the $B_{2g}$ channel is strong (Figs.\,\ref{Fig1}B and \ref{Fig2}); it is composed of several spectral features:

1) A low-energy QEP. 
The intensity of the QEP is weak at high temperatures. 
Upon cooling, the QEP softens from about 100 down to a few tens of~cm$^{-1}$, gains intensity, reaches its maximum intensity just above $T_S$, and then gradually loses its intensity below $T_S$ (the blue component in Fig.\,\ref{Fig2}B). 
In the SC phase, the QEP acquires coherence and undergoes a metamorphosis into an in-gap collective mode (Fig.\,\ref{Fig1}B, \textit{Inset}) similar to several other FeSC superconductors~\cite{Hackl_PRL2013,Chubukov_PRB2016,Thorsmolle_PRB2016,Gallais_PRL2016,SWu_PRB2017}. 

2) A broad electronic continuum extending beyond 2,000\,cm$^{-1}$ with the intensity peaking at about 450~cm$^{-1}$ and showing only weak dependence on temperature and doping (the yellow component in Fig.\,\ref{Fig2}B1-B5 and Fig.\,\ref{Fig3}A and B).

3) Below $T_S$, a significant pseudogap-like suppression develops at frequencies below 400~cm$^{-1}$ (Figs.\,\ref{Fig2} and \ref{Fig3}A). 

4) On approach to $T_S(x)$ an additional sharp low-frequency feature appears that is most pronounced for the alloys with high sulfur concentration, shown in violet in Fig.\,\ref{Fig2} B1-B5. 
The mode could be attributed to the lattice dynamics above $T_S(x)$ governed by back and forth fluctuation between two short-range nematic distortion domains~\cite{Billinge2019,Birgeneau2019} which break the symmetry in the opposite sense in the presence of local defects due to sulfur substitution, a feature typical for displacive structural phase transitions~\cite{halperin1976}. 

5) An additional intensity which develops below $T_S(x)$ at the lowest frequencies and is related to coupling between QEP response and acoustic lattice modes in the presence of a quasiperiodic array of the structural domain walls that appear in twinned crystals~\cite{Mai2021}.

6) A weak feature at about 190~cm$^{-1}$ that is related to interband transition between occupied $\beta$ and unoccupied $\alpha$ bands~\cite{Hashimoto2020}.

In Fig.\,\ref{Fig2}, we show doping dependence of $XY$ Raman response for FeSe$_{1-x}$S$_x$ with five sulfur concentrations. 
For $x$<0.16 alloys the tetragonal-to-orthorhombic structural phase transition temperature $T_S(x)$~\cite{Hosoi_pnas2016} is marked in Fig.\,\ref{Fig2}\,A1-A4. 
For all concentrations $x$\,<\,0.16, we observe an enhancement and critical softening of the QEP upon cooling toward $T_S$. 
Upon entering into the orthorhombic phase, intensity of the QEP diminishes and a pseudogap-like suppression develops at low frequencies. 
At the lowest temperature, a full gap suppression appears in the continuum for all samples with substitution concentrations $x$\,<\,0.15.
For $x=$\,0.15, some residual scattering intensity remains in the gap.
The energy of gap-like suppression onset (Fig.\,\ref{Fig2}\,B1-B4) appears to be close to the $d_{xz/yz}$ orbital splitting near the electron pocket in the nematic phase, as was reported by ARPES (Fig.\,\ref{Fig2}\,B1-B4)~\cite{Coldea_PRB2015,Yi_2019}. 
For the $x=$ 0.20 sample which remains tetragonal in the whole temperature range, no gap-like suppression is observed in the Raman spectra (Figs.\,\ref{Fig2}\,B5 and \ref{Fig3}B).
             
The peak in the broad continuum at about 450~cm$^{-1}$ appears at low temperatures for crystals with all sulfur compositions $x$. 
In Fig.\,\ref{Fig3} we show a comparison of $XY$-symmetry Raman response for pristine FeSe ($x=0$, $T_S = 88$~K) and heavily sulfur substituted ($x=0.20$) crystals. 
The 450-cm$^{-1}$ feature can be followed for both samples at all measured 
temperatures, in both tetragonal and orthorhombic phases. 
More importantly, for each given temperature the feature's line shape is quite similar for both samples: 
The only distinction between the data in tetragonal and orthorhombic phases is the pseudogap-like suppression which develops below $T_S$.  
Thus, this broad feature is not exclusive for the nematic phase 
(\textit{SI Appendix, Doping Dependence of the 450~cm$^{-1}$ Feature}).

\section*{Discussion}~~~
We model the $XY$-symmetry electronic Raman response containing the QEP and the 450\,cm$^{-1}$ features  by two main Raman oscillators with frequency-dependent self-energy (Figs.\,\ref{Fig2} B1-B5 and \ref{Fig3} )(\textit{SI Appendix, Data Fit}).   
In the tetragonal phase, the following form of self-energy provides the best data description:
\begin{equation}
\Sigma^{\prime\prime}(\omega, T)=m_0\omega+n_0 T, \; \; 
\text{if} \; T > T_{S}(x). 
\label{sigmatetr}
\end{equation}
The Fermi-liquid-like frequency dependence recovers in the low-temperature nematic phase, the green region in the phase diagram, Fig.\,\ref{Fig4}D, where the self-energy 
\begin{equation}
\Sigma^{\prime\prime}(\omega, T) = \left\{
\begin{array}{ll}
   m_0\omega^2/\omega_c(T,x) + n_0 T, & \text{if} \; \omega < \omega_c(T,x)\\
   m_0\omega+n_0 T, & \text{if} \; \omega > \omega_c(T,x)
\end{array}
\right.
\label{sigmanematic}
\end{equation}
represents the data best (Figs.\,\ref{Fig2} and \ref{Fig3}).   
Here $m_0 = 1.2$, $n_0 = 1.5$\,cm$^{-1}$/K, and $\omega_c(T,x)$ labels the cross-over frequency which evolves with temperature and sulfur doping, as shown in Fig.\,\ref{Fig4}C, similar to the onset energy of the pseudogap suppression. 
Consistent with this result, a low-temperature cross-over from quasi-$T$-linear to $T^2$ Fermi-liquid-like behavior was also reported for resistivity measurements $\rho(T,x)$ if $x < 0.16$ in the orthorhombic phase~\cite{Licciardello2019_Nature,Hussey_2019}. 

We note here that the appearance of strong low-frequency mode with the lattice involvement, especially for the $x=0.2$ sample (purple shading in the phase diagram, Fig.\,\ref{Fig4}D), is expected to quench the long-wavelength nematic fluctuations at low frequency~\cite{Oganesyan_2001,Maslov_PRL_2011,Paul_PRL_2017,Berg_2019}, giving rise to recovery of $T^2$-like resistivity with a significant residual value $\rho_0$, again consistent with the data in refs.\,\cite{Licciardello2019_Nature,Hussey_2019}.

Next, in Fig.\,\ref{Fig4} A1-A5 we plot temperature dependence of the static electronic Raman susceptibility for the QEP and the continuum contributions $\chi_{QEP}(0,T)$ and $\chi_C(0,T)$ that we derive from the spectra by Kramers-Kronig transformation:
\begin{equation}
\chi(0,T)=\frac{2}{\pi}\,P\int^{\omega_{uv}}_0\frac{\chi^{\prime\prime}(\omega,T)}{\omega}d\omega,
\end{equation}
where we choose high-energy cutoff $\omega_{uv}$ at 2,000\,cm$^{-1}$ (\textit{SI Appendix, Static Susceptibility}).
In contrast to mild temperature evolution of the continuum contribution $\chi_C(0,T)$, a critical enhancement upon cooling toward $T_S(x)$ is clearly observed for the QEP component of the static susceptibility. 
We fit the latter by an inverse power law 
\begin{equation}
1/\chi_{QEP}(0, T>T_S)=f(T) \propto \left[\frac{T-T_{QEP}(x)}{T_0}\right]^2+C,
\label{fT}
\end{equation}
(Fig.\,\ref{Fig4} B1-B5), where critical temperature $T_{QEP}(x)$ is shown in Fig.\,\ref{Fig4}D and 
$T_0$ is an effective temperature about 220\,K.
Note that the non-Curie-Weiss form of susceptibility arises due to self-energy effects in the tetragonal phase.

%Fig5
%%%%%%%%%%%%%%%%%%%%%%%%%%%%%%%%%%%%
\begin{figure}[!]
\centering
\includegraphics[width=\columnwidth]{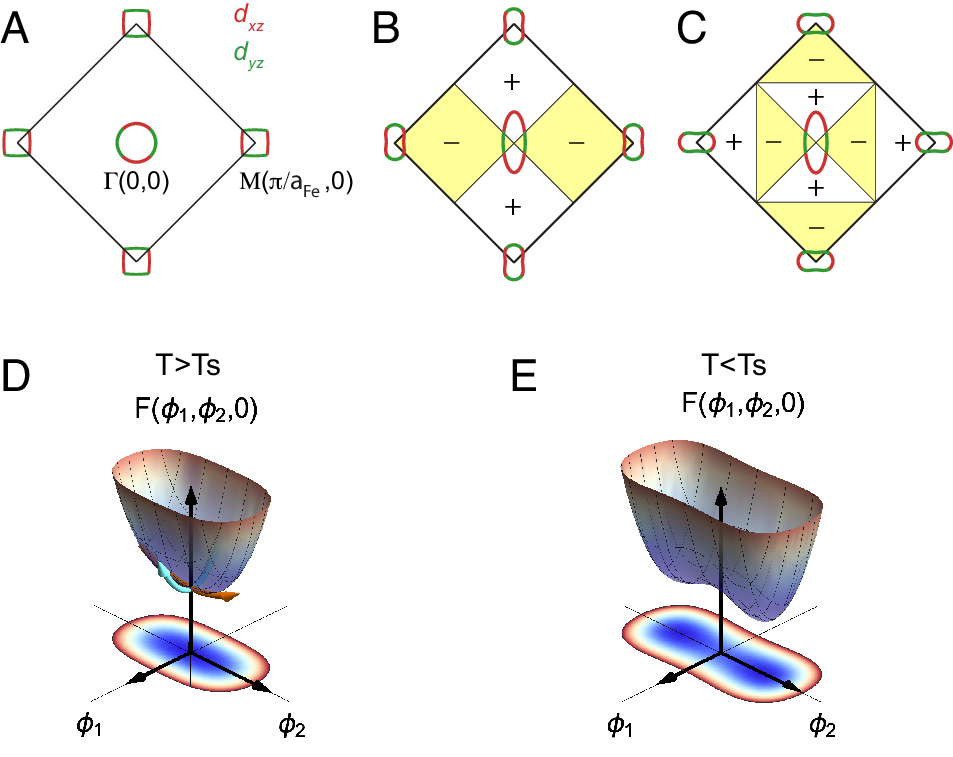}
\caption{\label{Fig5} (A) A minimal $d_{xz}$-$d_{yz}$ two orbital model consisting of a hole pocket at the $\Gamma$ point and an electron pocket at the M point for the high-temperature tetragonal phase. 
(B and C) Illustrations of $B_{2g}$ symmetry in-phase and antiphase Fermi surface distortions. 
(D and E) Ginzburg-Landau free energy for the tetragonal phase and the nematic phase. 
The orange and cyan arrows in D represent the oscillations of the order parameters in the high temperature phase that correspond to the QEP (feature 1) and the broad electronic continuum peaked at 450 cm$^{-1}$ (feature 2) in the Raman spectra, respectively.}
\end{figure}
%%%%%%%%%%%%%%%%%%%%%%%%%%%%%%%%%%%%%%  

The temperature dependence of the low-frequency QEP fluctuations has been meticulously studied for generic FeSC~\cite{Gallais_PRL2013,WLZhang_arxiv2014,Hackl_nphys2016,Massat_2016,Thorsmolle_PRB2016,Gallais2020}. 
The behavior arises from degeneracy of the partially filled iron 3$d_{xz}$ and 3$d_{yz}$ orbitals in the tetragonal phase~\cite{Vafek_PRB2013,JPHu_PRX2012,Weilu2018PRBRC}.
The QEP is related to overdamped dynamical charge oscillations at sub-terahertz frequencies, which give rise to a fluctuating charge quadrupole moment with an amplitude proportional to oscillating $d_{xz}/d_{yz}$ orbital charge imbalance $Q\propto n_{xz}-n_{yz}$, where $n_{xz/yz}$ is the orbital occupancy~\cite{WeiKu_PRL2009,Phillips_PRB_2009,Saito_PRB2010,Yamase_PRB2013,WLZhang_arxiv2014,Gallais_review,Kontani_PRL_2014,Thorsmolle_PRB2016,Chubukov_PRB2018}. 
Such excitations result in Pomeranchuk-like nematic dynamic deformation of the Fermi surface pockets with nodal lines in the $X/Y$ directions (see illustration of a snapshot in Fig.\,\ref{Fig5}C). 
These Pomeranchuk-like quadrupole fluctuations are strongly overdamped in the normal state leading to the QEP feature, while in the SC state, when low-energy relaxation is removed, the QEP feature transforms into a sharp in-gap collective mode~\cite{Thorsmolle_PRB2016,Gallais_PRL2016}. 
For pnictides, the temperature dependence of the bare static electronic susceptibility $\chi_{QEP}(0,T)$ often shows critical behavior on its own~\cite{Thorsmolle_PRB2016,Pomeranchuk_1958}, leading to a $d$-wave Pomeranchuk instability at enhanced temperature $T_S$ as a result of coupling to the lattice~\cite{Bohmer_2015,Gallais_review}. 
In the low-temperature orthorhombic phase, the fourfold rotational symmetry on the Fe site is broken and hence the degeneracy of Fe 3$d_{xz}$ and 3$d_{yz}$ is lifted, which causes rapid suppression of the low-energy fluctuations.

To explain the two components in FeSe$_{1-x}$S$_x$ spectra we construct a nominal model containing one hole FS pocket at the $\Gamma$ point and one electron pocket at the M point consisting only of $d_{xz}$ and $d_{yz}$ orbital characters (Fig.\,\ref{Fig5}). 
The two FS pockets give rise to two types of Pomeranchuk oscillations: in phase (Fig.\,\ref{Fig5}\,B) and anti-phase (Fig.\,\ref{Fig5}\,C). 
We define two order parameters 
$\phi_1$=$\phi_\Gamma$+$\phi_M$ and 
$\phi_2$=$\phi_\Gamma$-$\phi_M$, 
which depict these fluctuation phases, respectively.

For the tetragonal phase above $T_S(x)$, the fluctuations of the two order parameters around free-energy minima ($\phi_1$,$\phi_2$)=(0,0) give rise to the $XY$ symmetry Raman response (Fig.\,\ref{Fig5}D). 
The critical oscillations are in the direction with lowest free energy, they give rise to the QEP (feature 1), while the oscillation in the direction with the higher energy that is only weakly dependent on temperature and doping $x$ results in the broad electronic Raman continuum peaked at about 450\,cm$^{-1}$ (feature 2).

In the nematic phase below $T_S$ the system condenses into a state with minimal free-energy determined by the relation between intrapocket nematic interaction versus interpocket repulsion: 
If the former interaction prevails, the $\phi_1$ in-phase arrangement wins;
alternatively, if the interpocket repulsion is stronger than intrapocket nematic interaction, the $\phi_2$ antiphase arrangement is the ground state~\cite{Udina_FeSe_2019}.
The recent ARPES study indicates that in the nematic phase the $d_{xz}$ orbital dominates the FS pocket at the $\Gamma$ point while the $d_{yz}$ orbital dominates the pocket at the M point~\cite{Yi_2019}. 
Thus, the $\phi_2$ anti-phase FS distortion arrangement due to the interpocket repulsion is the prevailing order parameter~\cite{Udina_FeSe_2019} (see Fig.~\ref{Fig5}C). 
We also note that such configuration is antagonistic to a spin-density-wave order.  

Hence, the $\phi_2$ antiphase FS distortion fluctuations are responsible for both the critical quasi-elastic scattering in the tetragonal phase~\cite{Udina_FeSe_2019} and the in-gap collective mode in the SC phase~\cite{Thorsmolle_PRB2016}. 
Then, the $\phi_1$-like in-phase fluctuations of the FS distortion give rise to the broad high-energy Raman continuum. 
When the $d_{xz}/d_{yz}$ orbitals split in the ordered state below $T_S$~\cite{Borisenko_SciRep2017,Chubukov_PRB2017,Watson_PRB2017,Coldea_review2017,Yi_2019}, stiffness of the dominant $\phi_2$ order parameter suppresses the $\phi_1$-like quadrupole fluctuations at the frequencies below the orbital splitting energy, which naturally explains the recovery of Fermi-liquid regime seen as the appearance of a pseudogap in the $XY$-symmetry Raman response (Fig.\,\ref{Fig2} and \ref{Fig3}A). 

In Fig.\,\ref{Fig5}\,D and E we show the free energy as a function of $\phi_1$ and $\phi_2$ for above and below $T_S$ phases. 
Above $T_S$, the fluctuations of both order parameters $\phi_2$ and $\phi_1$ contribute to the Raman response consisting of the QEP at low energy and the high-energy continuum at around 450 cm$^{-1}$, respectively. 
Below $T_S$, the $\phi_2$ becomes the dominant order, while the $\phi_1$-like fluctuations are suppressed at low energies. 

Finally, we consider coupling of the critical $\phi_2$ order parameter to the orthorhombic lattice strain $\epsilon$ by constructing a model free energy of the system~\cite{Bohmer_2015}: 
\begin{equation}
F(\phi_2,\epsilon)=\frac{1}{2} f(T) \phi_2^2 + \frac{1}{4}\beta_2\phi_2^4 + \frac{C_{66,0}}{2} \epsilon^2 -\lambda \phi_2\epsilon,  
\end{equation}
where $f(T)$ is inverse bare electronic nematic susceptibility, $\lambda$ is bilinear electron-lattice coupling between $\phi_2$ and the orthorhombic lattice distortion $\epsilon$ with $B_{2g}$ symmetry, and $C_{66,0}$ is lattice bare shear modulus. 
Here we neglect the contributions due to subdominant electronic order $\phi_1$. 

In the following, we deduce the nematic susceptibility.
To consider the coupling of the electronic order parameter $\phi_2$ to the $XY$-symmetry Raman field \textbf{$A$} exerted by the $XY$ polarized incident and scattered light, we add to free energy an interaction term $-\gamma \phi_2 A$, here $\gamma$ is the interaction constant.  
Then, the nematic susceptibility can be expressed as response of electronic order parameter $\phi_2$ to perturbation $A$:
\begin{equation}
\chi_{nem}(T) = \frac{\partial \phi_2}{\partial A} = \left\{ 
\begin{array}{ll}
\frac{\gamma}{[f(T)-f(T_S)]}, & \text{if} \;  T > T_S\\
\frac{\gamma}{2 [f(T_S)-f(T)]}, & \text{if} \;  T_{QEP} < T < T_S
\end{array}
\right.,
\label{totalchi}
\end{equation}
where $T_S$ is defined by equation $f(T_S) = \lambda^2/C_{66,0}$. 
One can see from Fig.\,\ref{Fig4}B1-B4 that $f(T_S)$ and therefore the electron-lattice coupling $\lambda$ only mildly depend on $x$. 
Thus, the collapse of $T_S(x)$ at approach to $x_{cr}$ is primarily caused by suppression of $T_{QEP}(x)$. 
We also note that superconducting $T_c(x)$ is not enhanced in vicinity of $x_{cr}$, instead, both $T_{QEP}(x)$ and $T_c(x)$ have a weak enhancement in the middle of the nematic phase (Fig.\,\ref{Fig4}D).

In summary, we have demonstrated that polarization-resolved Raman spectroscopy provides detailed information on non-Fermi-liquid quadrupolar charge dynamics. 
In application to the nonmagnetic FeSe$_{1-x}$S$_x$ superconductors, we argue that the intense $XY$-symmetry Raman continuum of excitations in the high-temperature tetragonal phase arises due to non-Fermi-liquid dynamics governed by Pomeranchuk fluctuations and that these fluctuations are suppressed in the symmetry-broken orthorhombic phase enabling the recovery of Fermi liquid properties, in agreement with the transport studies~\cite{Licciardello2019_Nature}. 
We further show that while the tetragonal-to-orthorhombic phase transition is driven by the Pomeranchuk fluctuation soft mode, coupling to the lattice significantly enhances the nematic transition temperature.

\section*{Materials and Methods}
~~~FeSe$_{1-x}$S$_x$ ($x$ = 0, 0.04, 0.08, 0.15 and 0.2) single crystals were grown by the chemical vapor transport technique as described in~ref.\cite{Hosoi_pnas2016}. 
Substitution of sulfur for selenium acts as negative pressure, which suppresses $T_S$ while the system remains nonmagnetic, and superconductivity remains robust~\cite{Matsuura2017,Hosoi_pnas2016,Hanaguri_2017}.
Strain-free crystals were cleaved in a nitrogen atmosphere and positioned in a continuous-flow optical cryostat.

Polarization-resolved Raman spectra were acquired in a quasi-backscattering geometry from the $ab$ surface. We used 2.6-eV excitation from a Kr$^+$ laser. 
The laser power was kept below 10\,mW for most measurements and less than 2\,mW for the measurements in the SC state. 
The laser heating $\approx$ 1\,K/mW was estimated by the appearance of the stripe pattern on the crystal surface at $T_S$~\cite{Hackl_nphys2016}. 
The Raman scattering signal was analyzed by a custom triple-grating spectrometer and the data were corrected for the spectral response of the spectrometer.

Raman scattering spectra were acquired in three polarization
configurations ($\mu\nu$ = $XY$, $ab$ and $aa$) to separate
excitations in distinct symmetry channels: $B_{1g}=ab$, $B_{2g}=XY$,
and $A_{1g}=aa(bb)-XY$ )(\textit{SI Appendix, Background Subtraction} and \textit{SI Appendix, Doping Dependence of Phonon Spectra}).

The spectroscopic work at Rutgers (W.Z., S.W., and G.B.) was supported by NSF Grant DMR-1709161. The sample growth and characterization work in Japan was supported by Grants-in-Aid for Scientific Research (no. JP18H01177, no. JP18H05227, and no. JP19H00649), Innovative Area ``Quantum Liquid Crystals'' (no. JP19H05824) from the Japan Society for the Promotion of Science, and by CREST (no. JPMJCR19T5) from Japan Science and Technology. The work at National Institute of Chemical Physics and Biophysics (NICPB) was supported by the European Research Council under Grant 885413.

\onecolumngrid
\newpage

\setcounter{figure}{0}
 \renewcommand{\thefigure}{S\arabic{figure}}%
\begin{center}
\Large{\textbf{Supplementary Information for \\
Quadrupolar charge dynamics in the nonmagnetic FeSe$_{1-x}$S$_x$ superconductors}}
\end{center}

\appendix

\begin{flushleft}
\textbf{\large{Background subtraction}}
\end{flushleft}

~~~The imaginary part of the Raman susceptibility $\chi^{\prime\prime}_{\mu\nu}(\omega, T)$ is calculated from the total secondary emission intensity $I_{\mu\nu}(\omega,T)=[1+n(\omega,T)]\chi_{\mu\nu}^{\prime\prime}(\omega, T)+I_{lumi}$,
where $\mu(\nu)$ denotes the polarization of the incident and scattered light, $[1+n(\omega,T)]=[1-exp(-h\omega/k_BT)]^{-1}$ is the Bose distribution function for Stokes Raman scattering and $I_{lumi}$ is the luminescence background. 
The scattering intensity has been corrected for the system response and normalized by the incident laser power and the acquisition time. 

\begin{figure}[htbp]
\centering
\includegraphics[width=0.6\columnwidth]{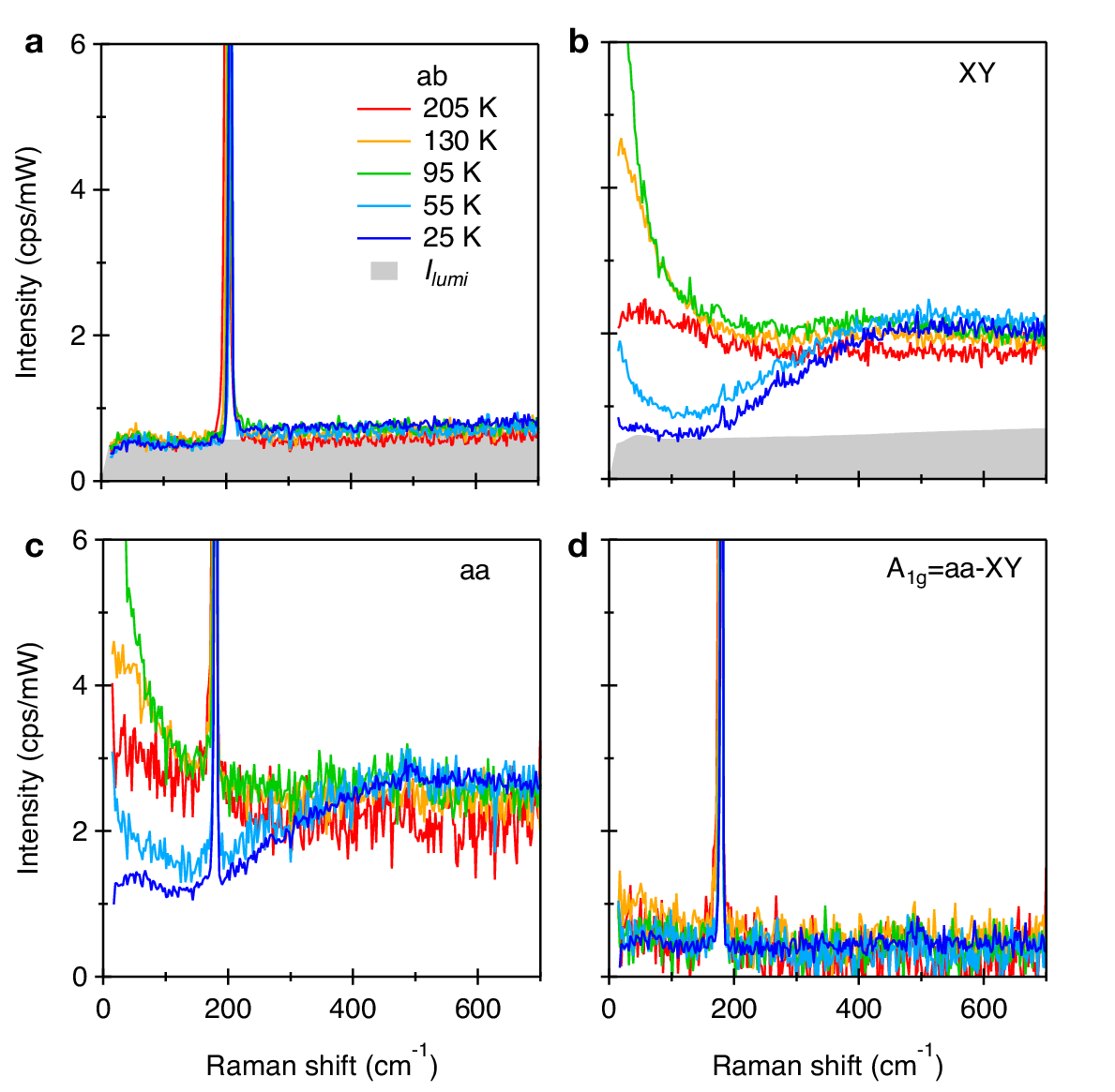}
\caption{\label{FigS1} (a-c) The secondary emission intensity for the
$ab$, $XY$ and $aa$ scattering geometries for undoped FeSe.
The luminescence background for the $ab$ and $XY$
scattering geometries is calculated from the emission continuum in
the $ab$ geometry. 
(d) $A_{1g}$ symmetry scattering intensity
calculated as the difference between secondary emission intensity in 
the $aa$ and $XY$ geometries.}
\end{figure}

Raman scattering spectra were acquired in three polarization
configurations ($\mu\nu$ = $XY$, $ab$ and $aa$) to separate
excitations in distinct symmetry channels ($B_{1g}=ab$, $B_{2g}=XY$,
and $A_{1g}=aa(bb)-XY$). In Figs.~\ref{FigS1}a and b we show the
secondary emission intensity for the $ab$ and $XY$ geometries at
various temperatures for undoped FeSe. 

The $ab$ geometry scattering continuum is almost independent of temperature; we attribute it to mainly luminescence background. 
Assuming that the luminescence is unpolarized (same for the $ab$ and $XY$ geometries), we
calculate the Raman response in the $ab$ and $XY$ scattering geometries with a temperature independent background estimated from the lowest $ab$ geometry scattering continuum ($B_{1g}$
phonon subtracted), as shown by the gray shade in Figs.~\ref{FigS1}a and b.

The $A_{1g}$ symmetry scattering intensity is calculated by subtracting the $XY$ symmetry secondary emission intensity from $aa$, as shown in Figs.~\ref{FigS1}c and d.

%%%%%%%%%%%%%%%%%%%%%%%%%%%%%%%%%%%%
%%%%%%%%%%%%%%%%%%%%%%%%%%%%%%%%%%%%%%

\newpage

\begin{flushleft}
\textbf{\large{Doping dependence of phonon spectra}}
\end{flushleft}

\begin{figure}[htbp]
\centering
\includegraphics[width=0.6\columnwidth]{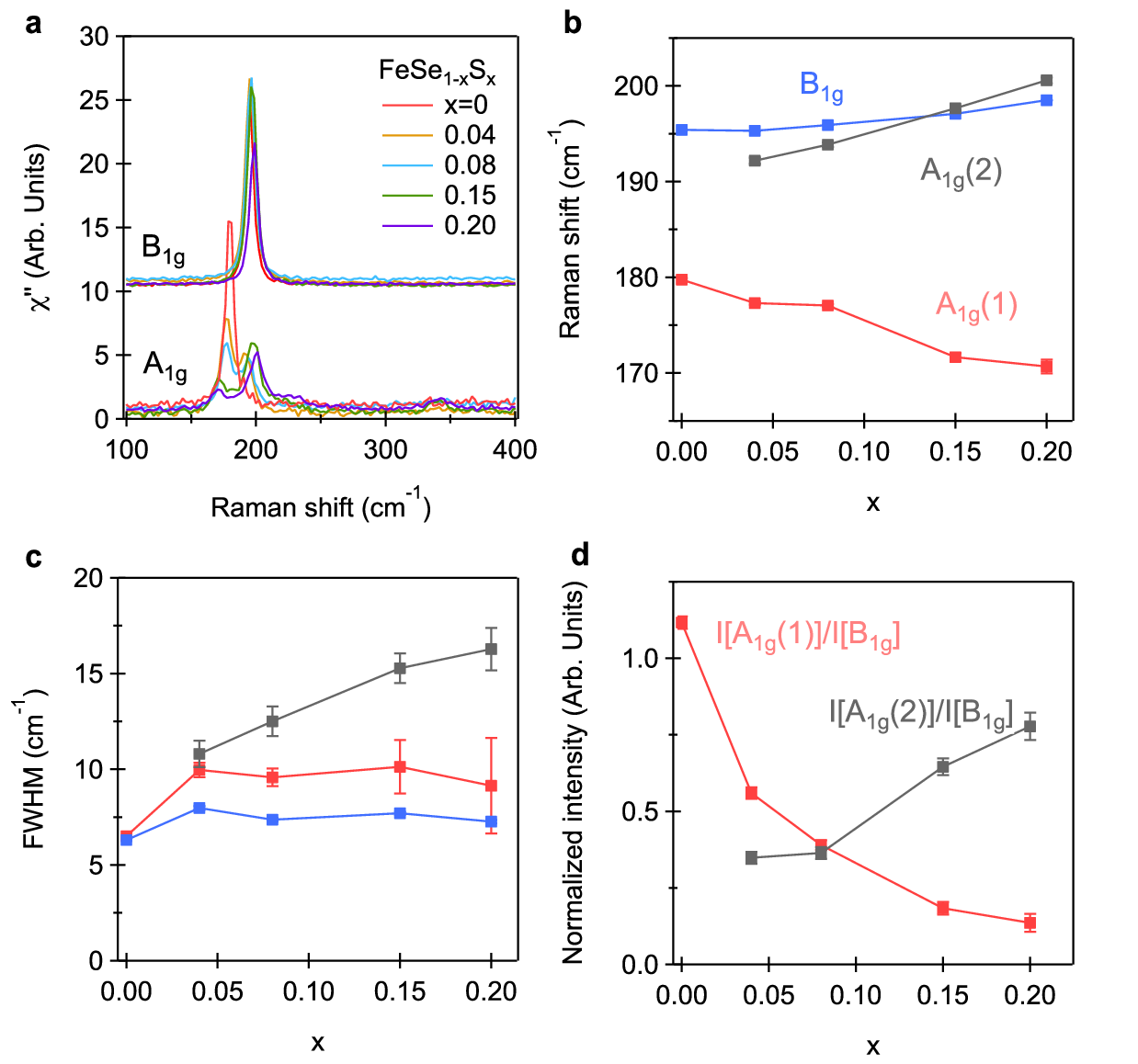}
\caption{\label{FigS2} (a) $A_{1g}$ and $B_{1g}$ symmetry phonon
spectra at the room temperature. 
The $B_{1g}$ spectra are offset in the
vertical direction. 
(b-c) Doping dependence of the energy and line width of the $B_{1g}$ and two $A_{1g}$ phonons. 
(d) The intensity (integrated area) of the two $A_{1g}$ phonons normalized to the intensity of $B_{1g}$ phonon.
%Error bars in (b) are determined by the instrument energy resolution 2.3~$cm^{-1}$. 
Error bars in (b)-(d) are the fitting standard errors.}
\end{figure}

~~~We observe two Raman active phonon modes from the $ab$-surface of undoped FeSe in quasi-back-scattering geometry at room temperature: 
an $A_{1g}$ phonon at 180~$cm^{-1}$ associated with Se vibrations and a $B_{1g}$ phonon at 195~$cm^{-1}$ associated with Fe vibrations~\cite{Gnezdilov_PRB2013}.  
In Fig.\,\ref{FigS2} we show the $A_{1g}$ and $B_{1g}$ phonon spectra at room temperature as well as  evaluation of the lineshape parameters (derived by Lorentzian fits) with increased sulfur concentration $x$. 
The normalized intensities shown in Fig.\,\ref{FigS2}(d) are the integrated areas of the $A_{1g}$ phonon normalized to the $B_{1g}$ phonon for the corresponding sulfur concentration $x$.

With the sulfur substitution $x$ the energy and line width of the $B_{1g}$ phonon changes insignificantly, while the $A_{1g}$ phonon gradually softens and loses intensity, while a new $A_{1g}$(2) phonon mode appears at around 193~$cm^{-1}$. 
The mode's frequency linearly hardens with $x$, thus it can be used for the sulfur content calibration. 
The normalized intensity of the $A_{1g}$ (2) phonon also increases with $x$. 
Such behavior is commonly observed for alloys in which the frequencies of the same phonon mode in the two end-point materials differ substantially. 
Importantly, we note that modes remain sharp with increased sulfur concentration, indicating high uniformity of the alloy crystals.

%%%%%%%%%%%%%%%%%%%%%%%%%%%%%%%%%%%%
%%%%%%%%%%%%%%%%%%%%%%%%%%%%%%%%%%%%%%

\newpage

\begin{flushleft}
\textbf{\large{Doping dependence of the 450 cm$^{-1}$ feature}}
\end{flushleft}

~~~A broad feature peaked at 450~cm$^{-1}$ is observed in the XY symmetry spectra for all studied FeSe$_{1-x}$S$_{x}$ alloys ($x =$ 0, \ldots , 0.2).  
Similar data were reported for undoped FeSe in
the prior literatures, however, inconsistent interpretations were
offered by different authors: 
in the Ref.~\cite{Massat_2016} the 450~cm$^{-1}$ feature was 
interpreted as nematic
response of ill-defined quasiparticles in a bad metal, while in 
Ref.~\cite{Hackl_Arxiv2017} the feature was interpreted as two-magnon
excitation whose intensity abruptly increase below $T_S$. 

\begin{figure}[htbp]
\centering
\includegraphics[width=0.35\columnwidth]{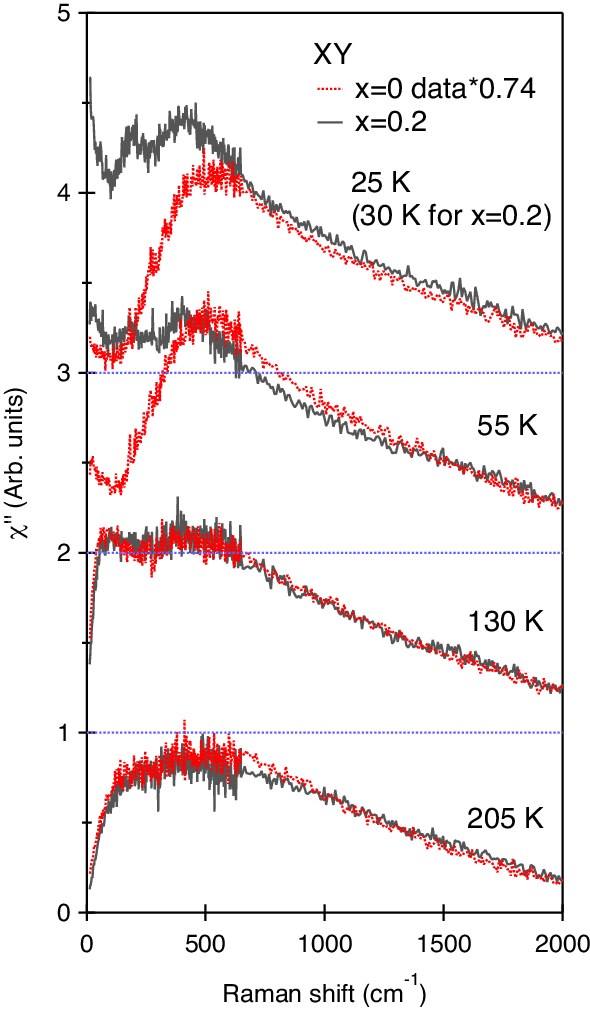}
\caption{\label{FigS3} Temperature evolution of $XY$ symmetry Raman
response for FeSe and FeSe$_{0.8}$S$_{0.2}$. 
The spectra are offset for clarity.}
\end{figure}

In Fig.\,2 (main text) and Fig.\,\ref{FigS3}, we directly compare temperature evolution of the $XY$ symmetry Raman response for undoped ($x=0$, $T_S$=88~K) and heavily sulfur doped ($x=0.2$, always tetragonal) crystals.

% at 205, 130, 55 and 25 (30)~K. 
% For $x=0$ sample, $T_S$=88 K. The $x$=0.2 sample is
% tetragonal for the whole temperature range. 
The 450 cm$^{-1}$ feature common for both samples at all measured temperatures, in both tetragonal and orthorhombic phases. 
This implies that this feature is only weakly effected by the nematic transition and that it remains vibrant when the alloy is doped further away from the magnetic phase~\cite{Matsuura2017}. 

More importantly, as shown in Fig.~\ref{FigS3}, for each given temperature, the lineshape is quite similar for both samples: the only distinction between the tetragonal and orthorhombic phases is the gap-like suppression in the broad continuum, which evolves as the nematic order parameter.
% (see also Fig.~\ref{FigS5}). 
This implies that sharpening of the 450~cm$^{-1}$ feature upon cooling is only due to temperature. 
% %
% \marginpar{\color{red} \tiny 
% A positive constructive conclusion is missing for this 
% important paragraph}
% %

Thus, this feature is not exclusive to the nematic phase. 
The data we present in Fig.~\ref{FigS3} neither supports the quasi-particle nematic response interpretation proposed in the Ref.~\cite{Massat_2016} nor the magnetic scattering interpretation proposed in the Ref.~\cite{Hackl_Arxiv2017}. 
Instead, we relate this mode to high energy in-phase nematic fluctuations at the $\Gamma$ and M points depict in Fig.\,5(b) of the main text.

%%%%%%%%%%%%%%%%%%%%%%%%%%%%%%%%%%%%

%%%%%%%%%%%%%%%%%%%%%%%%%%%%%%%%%%%%%%

\newpage

\begin{flushleft}
\textbf{\large{Data Fit}}
\end{flushleft}

We fit data for the whole temperature and sulfur doping range with a sum of electronic and lattice related  Raman oscillators. 
The electronic oscillators contain three terms:  
\begin{equation}
\label{Eqchi}
 \chi^{\prime\prime}(\omega,T,x) = \chi_1^{\prime\prime}[\omega,\Sigma^{\prime\prime}(\omega,T,x)] + \chi_2^{\prime\prime}[\omega,\Sigma^{\prime\prime}(\omega, T,x)] + \chi_3^{\prime\prime}(\omega,T,x) + \chi_4^{\prime\prime}(\omega,T,x)
 \end{equation}
where the first term describes the low-energy quasi-elastic scattering peak, the second term describes the high-energy broad electronic continuum peaked at about 450~cm$^{-1}$, the third term at about 190~cm$^{-1}$ is due to an interband transition between occupied $\beta$ and unoccupied $\alpha$ bands~\cite{Hashimoto2020}, and the fourth term is due to low-frequency lattice dynamics above $T_S(x)$ governed by back and forth fluctuation between two short-range nematic distortion domains~\cite{Billinge2019,Birgeneau2019} 
which break the symmetry in opposite sense in the presence of local defects due to sulfur substitution, a feature typical for displacive structural phase transitions~\cite{halperin1976}. 
Here $\Sigma^{\prime\prime}(\omega, T)$ is the imaginary self-energy.

\begin{flushleft}
\textbf{A. The self-energy}
\end{flushleft}

For the best fit we use following forms of self-energy:

\begin{enumerate}
\item In the tetragonal phase ($T>T_S$), 
\begin{equation}
\label{EqSISE1}
\Sigma^{\prime\prime}(\omega, T)=m_0\omega+n_0 T. 
\end{equation}
\item In the nematic phase ($T<T_S$), 
\begin{equation}
\label{EqSISE2}
\Sigma^{\prime\prime}(\omega, T, x) = \left\{
\begin{array}{ll}
   m_0\omega^2/\omega_c(T,x) + n_0 T, & \text{if} \; \omega < \omega_c(T,x)\\
   m_0\omega+n_0 T, & \text{if} \; \omega > \omega_c(T,x)
\end{array}
\right.
\end{equation}
where $\omega_c(T,x)$ is a crossover frequency which evolves similar to the gap suppression frequency, see Fig.\,4(c) in the main text. 
\end{enumerate}
For all sulfur compositions and all temperatures, $m_0 = 1.2$, $n_0 = 1.5$\,cm$^{-1}$/K.

\begin{flushleft}
\textbf{B. The low-energy quasi-elastic scattering peak (QEP)}
\end{flushleft}
  
\begin{flushleft}
\textit{\textbf{B.1. In the temperature regime $T>T_S$}}
\end{flushleft}

We describe the QEP contribution in high-temperature tetragonal phase with an overdamped Raman oscillator, the feature 1 shaded in blue in Fig.\,2(b1-b4):
 \begin{equation}
 \chi_1^{\prime\prime}[\omega,\Sigma^{\prime\prime}(\omega,T,x)]=\chi^{\prime\prime}_{QEP}[\omega,\Sigma^{\prime\prime}(\omega,T)]=T_{QEP}^2 Im\{[\omega+\omega_1+i\Sigma^{\prime\prime}(\omega,T)]^{-1}-[\omega-\omega_1+i\Sigma^{\prime\prime}(\omega, T)]^{-1}\}, 
\end{equation}
where non-Fermi-liquid like self-energy is given by Eq.\,[\ref{EqSISE1}] above. 
The resulted peak energy $\omega_{QEP}$ is shown in Fig.\,4(e).

\begin{flushleft}
\textit{\textbf{B.2. In the temperature regime $T<T_S$}}
\end{flushleft}

The Raman response in the orthorhombic phase below $T_S$ is described by the reminiscent QEP, the intensity of which is suppressed due to the iron 3$d_{xz}$ and 3$d_{yz}$ orbital splitting, that is coupled to $B_{2g}$-symmetry acoustic lattice modes in the presents of a quasi-periodic array of the structural domain walls that develop below $T_S(x)$ temperature~\cite{Mai2021}, the features 1 and 5 shaded in blue in Fig.\,2(b1-b4).
The coupled response can be described by Fano function, see Ref.\,\cite{Mai2021}:
\begin{equation}
\chi_1^{\prime\prime}[\omega,\Sigma^{\prime\prime}(\omega,T,x)]=Im\, [\,T^\dagger(G_0^{-1}-V)^{-1}T\,]
\label{Dyson}
\end{equation}
where $T^\dagger=(T_{QEP},T_A)$ is light coupling amplitude for the QEP and the acoustic mode. 
$T_A$ is set to be zero, because in the long wavelength limit direct Raman light field coupling to acoustics is weak. 
$G_0$= $\big(\begin{smallmatrix} G_{QEP}  & 0\\ 0 & G_{A}  \end{smallmatrix}\big)$ is a diagonal matrix in which $G_{QEP}$ and $G_{A}$ represent the bare Raman response for the QEP and the acoustic mode:
\begin{equation}
 G_{QEP}[\omega,\Sigma^{\prime\prime}(\omega,T,x)]=[\omega+\omega_1+i\Sigma^{\prime\prime}(\omega,T,x)]^{-1}-[\omega-\omega_1+i\Sigma^{\prime\prime}(\omega, T,x)]^{-1}
 \end{equation}
 \begin{equation}
G_A(\omega,x,q)=[\omega + c_s q + i r(x)q]^{-1}-[\omega - c_s q + i r(x)q]^{-1}, 
 \end{equation}
where $c_s=281$\,cm$^{-1}$\AA~is the unrenormalized sound velocity determined from inelastic neutron scattering data in Ref.\,\cite{Merritt_PRL2020}; 
$q=2\pi/d$ is the coupling wave vector due to quasi-periodic domain structure estimated from the domain size $d$ at about 4\,nm~\cite{Billinge2019,Birgeneau2019,Mai2021}. 
$V = \big(\begin{smallmatrix} 0 & \sqrt{q \omega_{QEP}(T_S) c_s/2}\\ \sqrt{q \omega_{QEP}(T_S) c_s/2} & 0 \end{smallmatrix}\big) $ 
in Eq.\,[\ref{Dyson}] is a real off-diagonal matrix that represents the interaction between the two modes, where $\omega_{QEP}(T_S)$ is the quasi-elastic peak energy at $T_S$, see Fig.\,4(e) in the main text~\cite{Mai2021}. 
We found that parameter $r(x)$ is linearly increasing with $x$ between 100 and 500\,cm$^{-1}$\AA.

\begin{flushleft}
\textbf{C. The broad electronic continuum}
\end{flushleft}
  
We describe the high-energy electronic continuum with another strongly overdamped Raman oscillator, the feature 2 shaded in yellow in Fig.\,2(b1-b4):
\begin{equation}
\chi_2^{\prime\prime}[\omega,\Sigma^{\prime\prime}(\omega,T,x)]=T_2^2 Im \{[\omega+\omega_2+i\Sigma^{\prime\prime}(\omega,T,x)]^{-1}-[\omega-\omega_2+i\Sigma^{\prime\prime}(\omega, T,x)]^{-1}\}
 \end{equation}
The mode's frequency $\omega_2$ is in between 800 and 1000\,cm$^{-1}$, its intensity changes weakly with temperature and doping in the whole temperature range. 
We note that due to strong relaxation the feature peaks at about 450\,cm$^{-1}$.

\begin{flushleft}
\textbf{D. The interband transition}
\end{flushleft}

We describe weak feature 6 at about 190 $cm^{-1}$ due (shaded in green in Fig.\,2(b1-b4)) to interband transition between occupied $\beta$ and unoccupied $\alpha$ bands by another Raman oscillator: 
\begin{equation}
 \chi_3^{\prime\prime}(\omega,T,x)=T_3^2 Im \{[\omega+\omega_3+i\Gamma_3]^{-1}-[\omega-\omega_3+i\Gamma_3]^{-1}\}.
 \end{equation}
$\omega_3$ and $\Gamma_3$ change only mildly with temperature and doping.

\begin{flushleft}
\textbf{E. The low-energy oscillator at above $T_S$}
\end{flushleft}

\begin{figure}[!htbp]
\centering
\includegraphics[width=0.65\columnwidth]{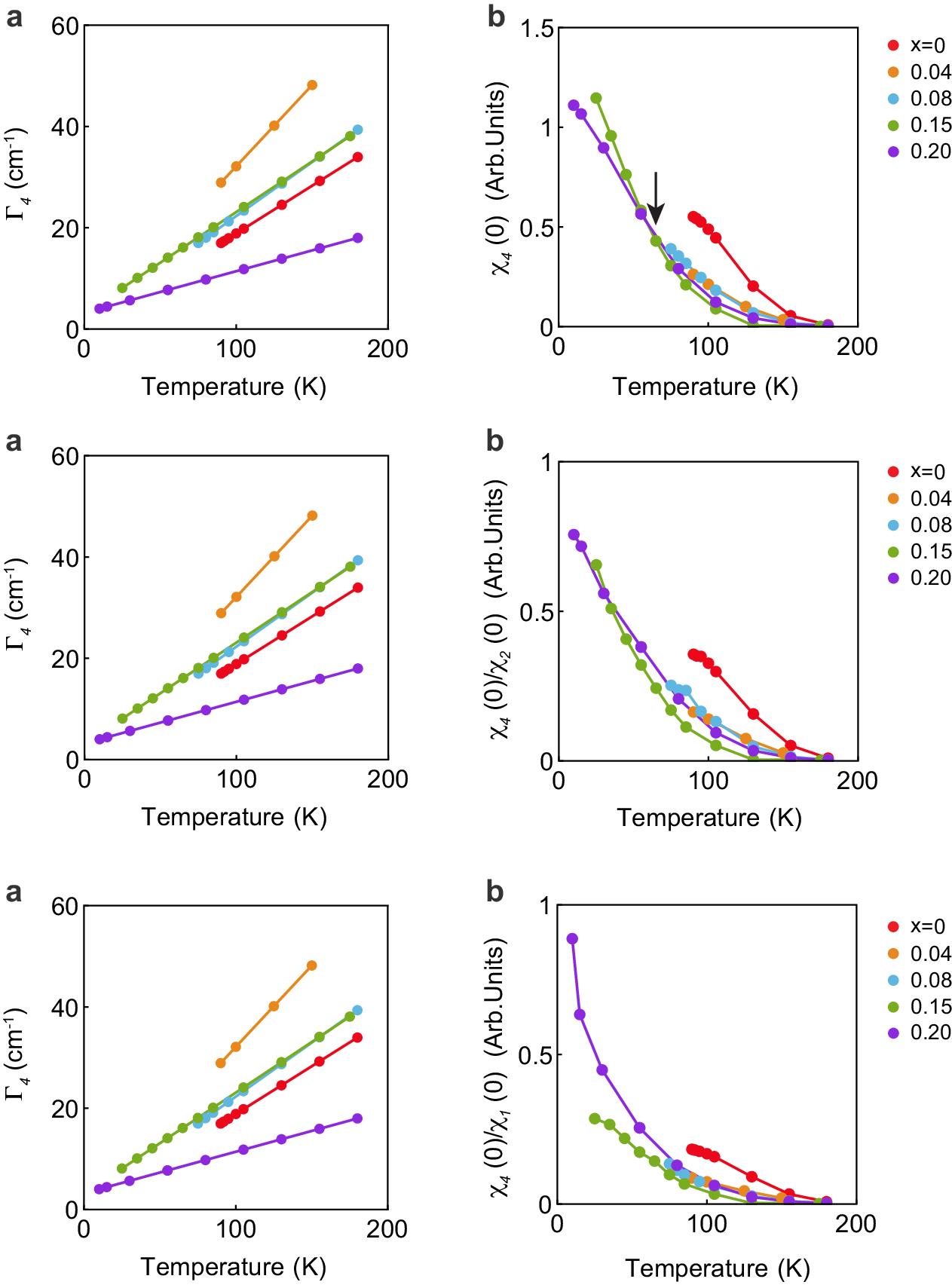}
\caption{\label{FigS11} 
Temperature dependence of (a) the scattering rate $\Gamma_4(T,x)$ and (b) the static Raman susceptibility $\chi_4(0,T,x)$ for the low-energy mode above $T_S$. 
The arrow in panel (b) indicates the crossover to purple region in the phase diagram in Fig.\,4(d), where $\chi_4(0,T,x)$ exceeds 0.5.}
\end{figure}

Finally, the contribution of the low-energy mode above $T_S$ (the feature 4 shaded in purple in Fig.\,2(b1-b4)) is described by another Raman oscillator with frequency at about 20\,cm$^{-1}$:
  \begin{equation}
\chi^{\prime\prime}_4(\omega,T,x)=T_4^2 Im\{[\omega+\omega_4+i\Gamma_4]^{-1}-[\omega-\omega_4+i\Gamma_4]^{-1}\}
 \end{equation}
Its scattering rate $\Gamma_4$ appears to increase linearly with temperature and with sulfur concentration $x$. 
The mode's contribution to static Raman susceptibility increases upon cooling towards $T_S$ and with the sulfur concentration $x$. 
The temperature dependence of the scattering rate and of the static Raman susceptibility for all $x$ are shown in Fig.~\ref{FigS11}.    
%%%%%%%%%%%%%%%%%%%%%%%%% 
%%%%%%%%%%%

\newpage

\begin{flushleft}
\textbf{\large{Static susceptibility}}
\end{flushleft}

\begin{figure}[!htbp]
\centering
\includegraphics[width=\columnwidth]{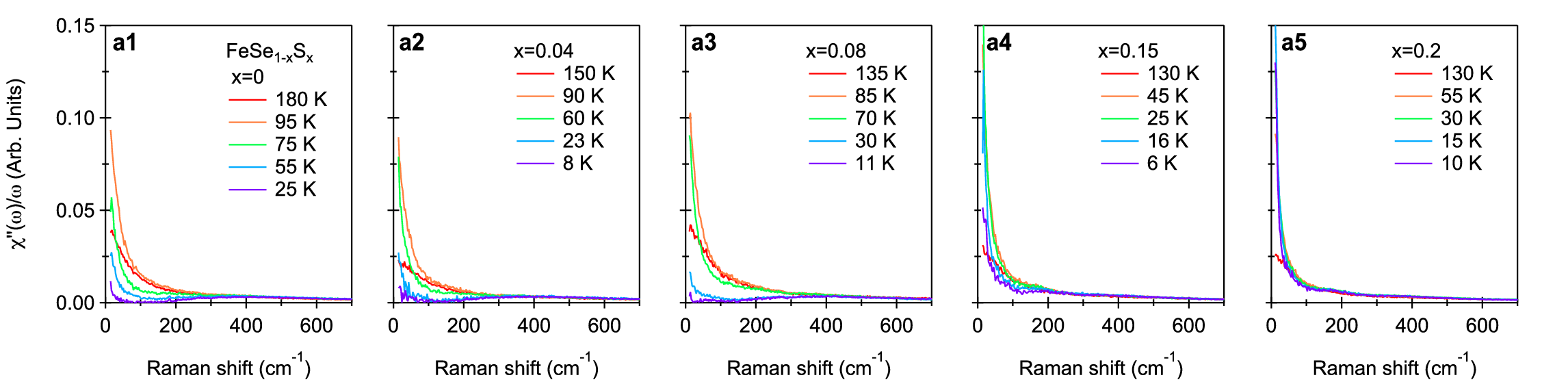}
\caption{\label{FigS6} $\chi^{\prime\prime}(\omega, T)/\omega$ at selected
temperatures for $x$ = 0, 0.04, 0.08, 0.15 and 0.2. }
\end{figure}

~~~The real part of the Raman susceptibility $\chi^{\prime}(\omega)$ can be derived from the imaginary part $\chi^{\prime\prime}(\omega)$ by Kramers-Kronig transformation. 
At $\omega=0$, $\chi^{\prime\prime}(\omega)$=0, the static susceptibility $\chi(0)$ is given by 
\begin{equation}
\chi(0,T,x)=\frac{2}{\pi}\,P\int^{\omega_{uv}}_0\frac{\chi^{\prime\prime}(\omega,T,x)}{\omega}d\omega,
\end{equation}
We calculate $\chi(0,T,x)$ from the fitting result of $\chi^{\prime\prime}(\omega,T)$ for the quasi-elastic peak, the high energy broad feature and the total Raman response, see Fig.\,4 (a1-a5, b1-b5). 
For the upper limit of the integration, we choose $\omega_{uv}$ = 2000~$cm^{-1}$ because $\chi^{\prime\prime}(\omega)/\omega$ becomes small for $\omega>500$\,cm$^{-1}$ (Fig.~\ref{FigS6}). 
For the response function below the low-energy measurement limit, we use extrapolation determined from the fitting parameter of $\chi^{\prime\prime}(\omega,T)$.

%%%%%%%%%%%%%%%%%%%%%%%%%%%%%%%%%%%%
%%%%%%%%%%%%%%%%%%%%%%%%%%%%%%%%%%%%%%
\newpage
\begin{flushleft}
\textbf{\large{Response in the superconducting phase }}
\end{flushleft}

%%%%%%%%%%%%%%%%%%%%%%%%%%%%%%%%%%%%
\begin{figure}[!htbp]
\centering
\includegraphics[width=0.8\columnwidth]{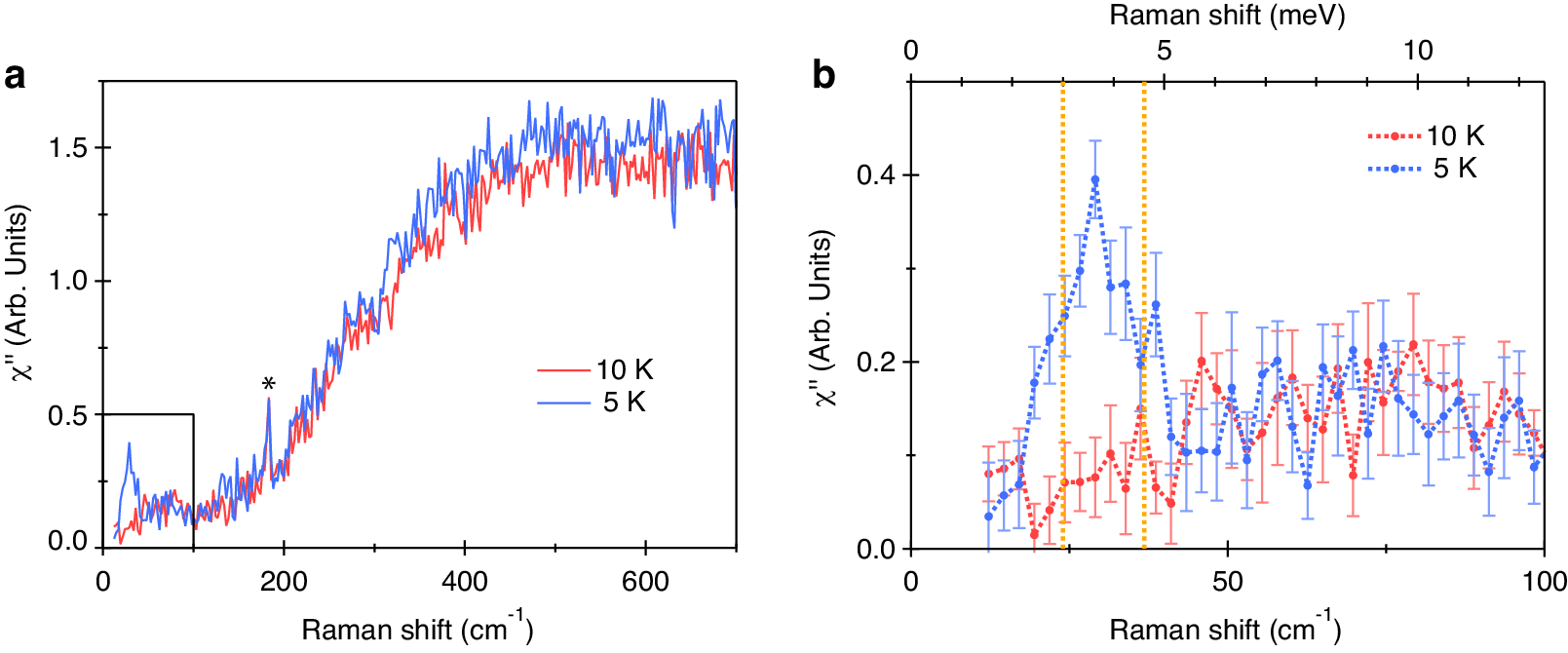}
\caption{\label{FigS7} 
(a) $XY$ symmetry Raman response in FeSe at 10\,K (normal state) and 5\,K (superconducting state). 
(b) Zoom in of the spectra in (a). 
The error bars are calculated  from the standard deviation. 
The magnitudes of the superconducting gaps 2$\Delta_{SC}$= 3 and 4.6\,meV defined by the scanning tunneling spectroscopy~\cite{JCDavis_Science2017} are shown by vertical dotted lines.
The mode at 183.5\,$cm^{-1}$ marked with an asterisk in panel (a) is the $A_g$ symmetry phonon mode. 
The phonon intensity appears in the $XY$ scattering geometry because the $A_{1g}$ and $B_{2g}$ symmetry channels merge when the high-temperature $D_{4h}$ group is reduced to the low-temperature $D_{2h}$ group~\cite{Wu_1712long}.}
\end{figure}
%%%%%%%%%%%%%%%%%%%%%%%%%%%%%%%%%%%%%%

~~~In Fig.\,\ref{FigS7} we show the $XY$ symmetry Raman response at 10\,K (normal state) and 5\,K (superconducting state).
In the superconducting state, the quasi-elastic scattering peak (QEP) is completely suppressed and a sharp symmetric collective mode at 29~cm$^{-1}$ (3.6 meV) appears. 
The mode's energy is between the two superconducting gap values 2$\Delta_{SC}=$3 and 4.6\,meV, as the gap values are determined by tunneling spectroscopy~\cite{JCDavis_Science2017}.

\bibliography{IBSC.bib}

\end{document}